\documentclass[10pt,twocolumn,reprint,amsmath,amssymb,aps,prl,showpacs,longbibliography,superscriptaddress]{revtex4-1}
\usepackage[latin9]{inputenc}
\usepackage{amsmath}
\usepackage{amssymb}
\usepackage{graphicx}
\usepackage{esint}

\makeatletter

\usepackage{amsfonts}
\usepackage{graphics}

\makeatother

\begin{document}
\title{Two-dimensional spectroscopic diagnosis of quantum coherence in Fermi
polarons}
\author{Jia Wang}
\affiliation{Centre for Quantum Technology Theory, Swinburne University of Technology,
Melbourne 3122, Australia}
\author{Hui Hu}
\affiliation{Centre for Quantum Technology Theory, Swinburne University of Technology,
Melbourne 3122, Australia}
\author{Xia-Ji Liu}
\affiliation{Centre for Quantum Technology Theory, Swinburne University of Technology,
Melbourne 3122, Australia}
\date{\today}
\begin{abstract}
We present a full microscopic many-body calculation of a recently-proposed
nonlinear two-dimensional spectroscopy for Fermi polarons, and show
that the quantum coherence between the attractive and repulsive polarons,
which has never been experimentally examined, can be unambiguously
revealed via quantum beats at the two off-diagonal crosspeaks in the
two-dimensional spectrum. We predict that particle-hole excitations
make the two crosspeaks asymmetric and lead to an additional side
peak near the diagonal repulsive polaron peak. Our simulated spectra
can be readily examined in future cold-atom experiments, where the
two-dimensional spectroscopy is to be implemented by using a Ramsey
interference sequence of rf pulses in the time domain. Our results
also provide a first-principle understanding of the recent two-dimensional
coherent spectroscopy of interacting excitons and trions in doped
monolayer transition metal dichalcogenides.
\end{abstract}
\maketitle
The polaron physics that describes the dynamics of a single impurity
interacting with a many-body environment is a long-standing problem
in modern physics \cite{Alexandrov2010}. The early study in 1933
by Lev Landua \cite{Landau1933} led to the cornerstone concept of
quasiparticles, which vividly characterizes the ability of the impurity
operating in its own, free-particle-like way in terms of a residue
$0<Z<1$. Over the next 70 years, sequent studies of the polaron problem
generated a number of celebrated ideas in many-body physics and condensed
matter physics, such as Kondo screening \cite{Hewson1993}, Anderson\textquoteright s
orthogonality catastrophe \cite{Anderson1967}, the x-ray Fermi edge
singularity \cite{Mahan1967,Roulet1969,Nozieres1969}, Nagaoka ferromagnetism
\cite{Nagaoka1966,Shastry1990,Basile1990} and the phase string effect
\cite{Sheng1996}.

\begin{figure}[b]
\centering{}\includegraphics[width=0.5\textwidth]{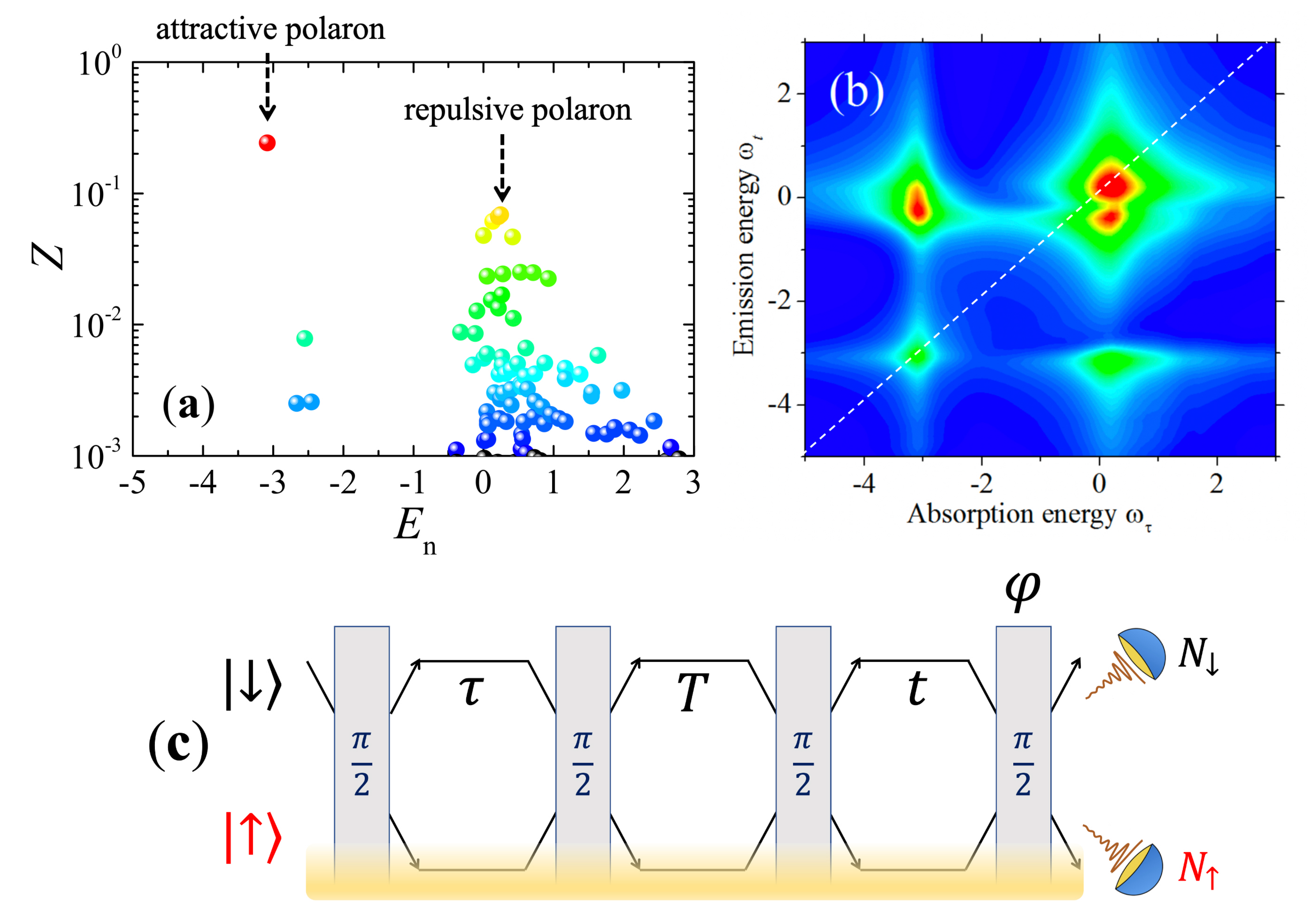}\caption{(a) An example of the residues at different many-body polaron states.
The ground-state attractive polaron and the excited repulsive polaron
at the energy $E_{A}$ and $E_{R}$ have been explicitly indicated.
(b) A typical 2DS spectrum of Fermi polarons at the mixing time $T=0$,
where the absorption energy ($\omega_{\tau}$) and emission energy
($\omega_{t}$) are obtained by Fourier transforming the time delays
$\tau$ and $t$, respectively. Two asymmetric crosspeaks at $(\omega_{\tau},\omega_{t})=(E_{A},E_{R})$
and $(E_{R},E_{A})$, off the diagonal direction (i.e., the dotted
line), reveal the coherence between the attractive and repulsive polarons.
(c) The 2DS pulse sequence in the time domain, defined by the time
delays ($\tau,T,t$). The impurity in the spin-up state interacts
with a background Fermi sea, as indicated by the shaded area. \label{fig:Sketch}}
\end{figure}

Over the past two decades, the polaron physics has received much more
intense interests, owing to the unprecedented controllability achieved
in ultracold atomic gases \cite{Bloch2008,Chin2010}. The dynamics
of an impurity atom immersed in a Fermi sea (Fermi polaron) \cite{Schirotzek2009,Zhang2012,Kohstall2012,Koschorreck2012,Cetina2016,Scazza2017}
or in a weakly interacting Bose condensate (Bose polaron) \cite{Hu2016,Jorgensen2016}
has now been systematically investigated in a quantitative manner
\cite{Massignan2014,Lan2014,Schmidt2018}, with precisely tunable
masses and interactions. A remarkable discovery in this context is
the observation of repulsive polaron \cite{Kohstall2012,Koschorreck2012,Scazza2017,Cui2010,Massignan2011},
which is a \emph{collection} of excited many-body states with non-negligible
residues close to a characteristic energy (i.e., repulsive polaron
energy), as illustrated in Fig. \ref{fig:Sketch}(a). The
repulsive polaron separates from the ground-state attractive polaron
by a spectral gap (i.e., dark continuum \cite{Goulko2016,Wang2022PRL,Wang2022PRA})
and can be quantitatively characterized in experiments by using injected
radio-frequency (rf) spectroscopy \cite{Scazza2017}. As an excited
quasiparticle, repulsive polaron is naturally anticipated to coherently
couple to attractive polaron, in the same way as an effective two-level
quantum system. Unfortunately, such a quantum coherence has never
been experimentally verified in cold-atom laboratories, by using either
Rabi-type or Ramsey-type interferometry \cite{Cetina2016,Schmidt2018,Knap2012}.

The purpose of this Letter is to present quantitative, experimentally
testable predictions on a novel nonlinear two-dimensional spectroscopy
(2DS) of Fermi polarons, which can provide an unambiguous spectroscopic
diagnosis of the quantum coherence between attractive and repulsive
polarons via quantum beats at two off-diagonal crosspeaks in the 2DS
spectrum, as shown in Fig. \ref{fig:Sketch}(b). This 2DS - implemented
by a sequence of Ramsey-type $\pi/2$ rf pulses as given in Fig. \ref{fig:Sketch}(c)
- was recently proposed by one of us in Ref. \cite{Wang2022PRX},
where exact quantum dynamics in the presence of an infinitely heavy
impurity has been considered. However, the immobile, heavy polaron
limit suffers from Anderson's orthogonality catastrophe that renders
Fermi polaron quasiparticles into power-law singularities \cite{Anderson1967,Knap2012}.
Thus, strictly speaking, it can only provide a qualitative understanding
for the 2DS of Ferm polarons. Here, such a difficulty is overcome
by a microscopic many-body calculation for a \emph{mobile} impurity
with finite mass. As a consequence, we are able to analytically clarify
that the Fermi sea shaking \cite{Schmidt2018,Knap2012}, in the form
of particle-hole excitations, makes the 2DS highly asymmetric. We
find that the Fermi sea shaking also introduces an interesting side
peak in the 2DS, slightly below the diagonal peak at the repulsive
polaron energy. 

It is worth noting that the 2DS is a cold-atom analogue of the well-known
two-dimensional coherent spectroscopy (2DCS) in condensed matter physics
\cite{Jonas2003,Li2006,Cho2008,Davis2008,Nardin2015}. The latter
has been widely used to reveal the many-body dynamics in semiconductors
\cite{Li2006,Nardin2015,Dey2016,Hao2016NatPhys,Hao2016NanoLett},
although its full potential is severely limited by the lack of theoretical
interpretation at the microscopic level \cite{Li2006,Tempelaar2019,Reichman2002}.
Interacting excitons and trions in doped monolayer transition metal
dichalcogenides (TMD) are intriguing examples \cite{Hao2016NanoLett,Tempelaar2019,Muir2022,Reichman2002}.
Remarkably, such systems have recently been understood as Fermi polarons
\cite{Sidler2017,Efimkin2017}, where excitons and trions can be precisely
re-interpreted as repulsive and attractive polarons, respectively.
Despite the different excitation schemes (i.e., the spin flip by rf
pulses in 2DS versus the exciton creation and annihilation by lasers
in 2DCS), we find that our simulated spectra provide an excellent
explanation to the experimental 2DCS of excitons and trions \cite{Hao2016NanoLett}.
Our results therefore present an exciting representative case, towards a full
ab initio understanding of the 2DCS in condensed matter.

\textbf{\textit{Model}}\textsl{.} The system under consideration consists
of a single spin-1/2 impurity (with creation operator $d_{\mathbf{k}\sigma}^{\dagger}$
for two hyperfine states $\sigma=\uparrow,\downarrow$) immersed in
a non-interacting Fermi bath (with creation operator $c_{\mathbf{k}}^{\dagger}$),
as described by the model Hamiltonian ($\hbar=1$),
\begin{equation}
\mathcal{H}_{\sigma}=\sum_{\mathbf{k}}\left[\epsilon_{\mathbf{k}}c_{\mathbf{k}}^{\dagger}c_{\mathbf{k}}+\left(\epsilon_{\mathbf{k}}^{I}+\omega_{s}\delta_{\sigma\uparrow}\right)d_{\mathbf{k}\sigma}^{\dagger}d_{\mathbf{k}\sigma}\right]+\mathcal{H}_{U}\delta_{\sigma\uparrow},
\end{equation}
when the impurity is the spin-$\sigma$ state. Here, $\epsilon_{\mathbf{k}}$
and $\epsilon_{\mathbf{k}}^{I}$ are respectively the kinetic energies
of the bath and impurity, $\omega_{s}$ denotes the energy difference
between the two spin states and is typically much larger than all
other energy scales in the problem, and $\delta_{\sigma\sigma'}$
is the usual Kronecker delta. The spin-up state of the impurity is
tuned by Feshbach resonance \cite{Chin2010} to be strongly interacting
with the Fermi bath, as described by the contact interaction Hamiltonian
$\mathcal{H}_{U}=U\sum_{\mathbf{qkp}}d_{\mathbf{k\uparrow}}^{\dagger}c_{\mathbf{q-k}}^{\dagger}c_{\mathbf{q}-\mathbf{p}}d_{\mathbf{p}\uparrow}$.
This gives rise to the many-body polaron states, as sketched in Fig.
\ref{fig:Sketch}(a). In contrast, the spin-down impurity state has
negligible interaction with the bath.

\textbf{\textit{Theory of 2DS}}\textsl{.} In the standard Ramsey interferometry
\cite{Cetina2016,Knap2012}, which involves only the first and the
final $\pi/2$ rf pulses in Fig. \ref{fig:Sketch}(c), the spin-down
impurity state acts a reference for phase evolution. The first pulse
turns the initially prepared spin-down state $\left|\downarrow\right\rangle $
into a superposition $(\left|\uparrow\right\rangle +\left|\downarrow\right\rangle )/\sqrt{2}$,
in which during the later evolution the spin-up state $\left|\uparrow\right\rangle $
acquires an additional phase due to the interaction with the Fermi
bath. This phase difference can be read out by applying the final
detection $\pi/2$ rf pulse and measuring the two occupation numbers
$N_{\uparrow}$ and $N_{\downarrow}$ \cite{Cetina2016,Knap2012}.
The resulting Ramsey response, given by the quantum average of the
Pauli matrix $\left\langle \sigma_{x}\right\rangle $, can reveal
the existence of both attractive and repulsive polarons \cite{Wang2022PRL,Wang2022PRA,Knap2012}.
In our 2DS measurement \cite{Wang2022PRX}, two more $\pi/2$ rf pulses
are utilized to explore the many-body evolution in the multidimensional
time domain and hence unfold quantum correlations between the two
polaron branches.

To show this, let us express the $\pi/2$ rf pulse in terms of the
operators $\hat{n}=\sum_{\mathbf{k}\sigma}d_{\mathbf{k}\sigma}^{\dagger}d_{\mathbf{k}\sigma}$
and $\hat{s}_{+}=\sum_{\mathbf{k}}d_{\mathbf{k}\uparrow}^{\dagger}d_{\mathbf{k}\downarrow}$,
i.e., $R_{\pi/2}=(\hat{n}+\hat{s}_{+}-\hat{s}_{-})/\sqrt{2}$, where
$\hat{s}_{-}\equiv\hat{s}_{+}^{\dagger}$. The time evolution between
two pulses is given by $\mathcal{U}(t')=e^{-i\mathcal{H}t'}$ for
$t'=\tau,T,t$ and $\mathcal{H}$ can be either $\mathcal{H}_{\uparrow}$
or $\mathcal{H}_{\downarrow}$ depending on the impurity state during
time evolution. Denoting the initial many-body state as $\left|\psi_{i}\right\rangle =d_{\mathbf{K}\downarrow}^{\dagger}\left|\textrm{FS}\right\rangle $,
where $\left|\textrm{FS}\right\rangle $ describes the Fermi sea at
zero temperature filled by particles with momentum $\left|\mathbf{k}\right|<k_{F}$
and the impurity is assumed to have a definite initial momentum $\mathbf{K}$,
the final state $\left|\psi_{f}\right\rangle $ before the last detection
pulse can be written as, 
\[
\left|\psi_{f}\right\rangle =\mathcal{U}\left(t\right)R_{\pi/2}\mathcal{U}\left(T\right)R_{\pi/2}\mathcal{U}\left(\tau\right)\frac{d_{\mathbf{K}\uparrow}^{\dagger}+d_{\mathbf{K}\downarrow}^{\dagger}}{\sqrt{2}}\otimes\left|\textrm{FS}\right\rangle .
\]
The measurement of the Pauli matrix $\sigma_{+}=\sigma_{x}+i\sigma_{y}=2\hat{s}_{+}$
at the detection stage then yields the 2DS response \cite{Wang2022PRX},
$\tilde{S}(\tau,T,t)=-2\left\langle \psi_{f}\left|\hat{s}_{+}\right|\psi_{f}\right\rangle $.

By inserting the expression of $R_{\pi/2}$ into $\left|\psi_{f}\right\rangle $,
it is straightforward to check that $\tilde{S}(\tau,T,t)$ has sixteen
different combinations \cite{Wang2022PRX}, each of which corresponds
to a pathway connecting the six unitary evolution operators $\mathcal{U}$
and has a different phase associated with the largest energy scale
$\omega_{s}$. As the rf pulse is in principle tuned in resonant with
$\omega_{s}$, we can take the rotating wave approximation and consider
only two dominant pathways \cite{Wang2022PRX}, $I_{i}(\tau,T,t)=\left\langle \textrm{FS}\right|d_{\mathbf{K}\downarrow}\hat{I}_{i}d_{\mathbf{K}\uparrow}^{\dagger}\left|\textrm{FS}\right\rangle $,
where 
\begin{eqnarray*}
\hat{I}_{1} & \equiv & e^{i\mathcal{H}_{\downarrow}\tau}\hat{s}_{-}e^{i\mathcal{H}_{\uparrow}T}\hat{n}e^{i\mathcal{H}_{\uparrow}t}\hat{s}_{+}e^{-i\mathcal{H}_{\downarrow}t}\hat{s}_{-}e^{-i\mathcal{H}_{\uparrow}T}\hat{n}e^{-i\mathcal{H}_{\uparrow}\tau},\\
\hat{I}_{2} & \equiv & e^{i\mathcal{H}_{\downarrow}\tau}\hat{n}e^{i\mathcal{H}_{\downarrow}T}\hat{s}_{-}e^{i\mathcal{H}_{\uparrow}t}\hat{s}_{+}e^{-i\mathcal{H}_{\downarrow}t}\hat{n}e^{-i\mathcal{H}_{\downarrow}T}\hat{s}_{-}e^{-i\mathcal{H}_{\uparrow}\tau}.
\end{eqnarray*}
There are also two pathways $I_{3}$ and $I_{4}$ that are of marginal
importance due to their fast-oscillating phase factor $e^{\pm i\omega_{s}T}$
at nonzero mixing time $T$ \cite{Wang2022PRX}. However, they can
easily be eliminated by a phase cycling procedure \cite{Wang2022PRX},
i.e., by considering another Ramsey sequence, in which after the $\tau$-delay
we take a $-\pi/2$ rf pulse instead of a $+\pi/2$ pulse. By denoting
the corresponding response as $\tilde{S}_{-}(\tau,T,t)$, we define
the phase cycling 2DS response that is of central interest \cite{Wang2022PRX},
\begin{equation}
\mathcal{S}\left(\tau,T,t\right)=\tilde{S}-\tilde{S}_{-}=\frac{I_{1}\left(\tau,T,t\right)+I_{2}\left(\tau,T,t\right)}{2}.
\end{equation}

In general, $I_{i}(\tau,T,t)$ ($i=1,2$) are extremely difficult
to calculate for an interacting many-body system. Nevertheless, for
Fermi polarons we can obtain the \emph{analytic} expressions of $I_{i}(\tau,T,t)$,
by taking the advantage that any ($n$-th) polaron state can be exactly
expressed through multiple-particle-hole excitations of the Fermi
sea \cite{Shastry1990,Basile1990,Chevy2006},
\begin{align*}
\left|n;\mathbf{k}\right\rangle  & =\sum_{\vec{\kappa}_{\nu}}\phi_{\vec{\kappa}_{\nu}}^{(n)}\left(\mathbf{k}\right)d_{\mathbf{k}-\mathbf{k}_{\vec{\kappa}_{\nu}}\uparrow}^{\dagger}\left|\vec{\kappa}_{\nu}\right\rangle ,
\end{align*}
where $\left|\vec{\kappa}_{\nu}\right\rangle =\prod_{i=1}^{\nu}c_{\mathbf{k}_{p}^{(i)}}^{\dagger}\prod_{i=1}^{\nu}c_{\mathbf{k}_{h}^{(i)}}\left|\textrm{FS}\right\rangle $
denotes $\nu$ particle-hole pairs excitations on top of a Fermi sea,
$\vec{\kappa}_{\nu}\equiv\{\mathbf{k}_{p}^{(1)},\mathbf{k}_{p}^{(2)},\cdots,\mathbf{k}_{p}^{(\nu)};\mathbf{k}_{h}^{(1)},\mathbf{k}_{h}^{(2)},\cdots,\mathbf{k}_{h}^{(\nu)}\}$
is a collective notation for the $\nu$ particle momenta ($\mathbf{k}_{p}^{(i)}>k_{F}$)
and hole momenta ($\mathbf{k}_{h}^{(i)}<k_{F}$), and therefore the
total momentum and energy of the particle-hole excitations are given
by $\mathbf{k}_{\vec{\kappa}_{\nu}}\equiv\sum_{i=1}^{\nu}[\mathbf{k}_{p}^{(i)}-\mathbf{k}_{h}^{(i)}]$
and $\epsilon_{\vec{\kappa}_{\nu}}=\sum_{i=1}^{\nu}[\epsilon_{\mathbf{k}_{p}^{(i)}}-\epsilon_{\mathbf{k}_{h}^{(i)}}]$,
respectively. At the leading order without particle-hole excitations,
we simply have $\left|\vec{\kappa}_{\nu=0}\right\rangle =\left|\textrm{FS}\right\rangle $
and $\phi_{\vec{\kappa}_{\nu=0}}^{(n)}(\mathbf{k})=\phi_{0}^{(n)}(\mathbf{k}).$
The energy of the ($n$-th) polaron state can be denoted as, $\mathcal{E}_{n}(\mathbf{k})=E_{\uparrow}^{(n)}(\mathbf{k})-(\epsilon_{\mathbf{k}}^{I}+\omega_{s}+E_{\textrm{FS}})$,
after the subtraction of the impurity energy ($\epsilon_{\mathbf{k}}^{I}+\omega_{s}$)
and the energy of the background Fermi sea ($E_{\text{FS}}$). On
the other hand, the many-body eigenstates in the case of the spin-down
impurity are much simpler and can be directly characterized by $\vec{\kappa}_{\nu}$,
i.e., $\left|\vec{\kappa}_{\nu};\mathbf{k}\right\rangle =d_{\mathbf{k}-\mathbf{k}_{\vec{\kappa}_{\nu}}\downarrow}^{\dagger}\left|\vec{\kappa}_{\nu}\right\rangle $.
The corresponding energy is given by, $\delta\mathcal{E}_{\vec{\kappa}_{\nu}}(\mathbf{k})=E_{\downarrow}^{(\vec{\kappa}_{\nu})}(\mathbf{k})-(\epsilon_{\mathbf{k}}^{I}+E_{\textrm{FS}})=\epsilon_{\vec{\kappa}_{\nu}}+\epsilon_{\mathbf{\mathbf{k}-\mathbf{k}_{\vec{\kappa}_{\nu}}}}^{I}-\epsilon_{\mathbf{k}}^{I}$,
which is a summation of recoil energy of the impurity and the Fermi
sea.

Let us now formally expand the time evolution operators as ($t'=\tau,T,t$),
\begin{eqnarray*}
e^{\pm i\mathcal{H}_{\uparrow}t'} & = & \sum_{n\mathbf{k}}e^{\pm iE_{\uparrow}^{(n)}(\mathbf{k})t'}\left|n;\mathbf{k}\right\rangle \left\langle n;\mathbf{k}\right|,\\
e^{\pm i\mathcal{H}_{\downarrow}t'} & = & \sum_{\vec{\kappa}_{\nu}\mathbf{k}}e^{\pm iE_{\downarrow}^{(\vec{\kappa}_{\nu})}(\mathbf{k})t'}\left|\vec{\kappa}_{\nu};\mathbf{k}\right\rangle \left\langle \vec{\kappa}_{\nu};\mathbf{k}\right|,
\end{eqnarray*}
and insert them into the expression of $I_{i}(\tau,T,t)$ ($i=1,2$).
By using the identities, such as $\left\langle n;\mathbf{q}\right|d_{\mathbf{k}\uparrow}^{\dagger}\left|\textrm{FS}\right\rangle =\phi_{0}^{(n)*}(\mathbf{k})\delta_{\mathbf{kq}}$
and $\left\langle \vec{\kappa}_{\nu};\mathbf{k}\right|\hat{s}_{-}\left|n;\mathbf{q}\right\rangle =\phi_{\vec{\kappa}_{\nu}}^{(n)}(\mathbf{k})\delta_{\mathbf{kq}}$,
after some straightforward algebra we find that,
\begin{eqnarray*}
I_{1} & = & \sum_{nm\vec{\kappa}_{\nu}}\Phi_{\vec{\kappa}_{\nu}}^{(nm)}e^{-i\mathcal{E}_{n}\tau}e^{-i\left(\mathcal{E}_{n}-\mathcal{E}_{m}\right)T}e^{i\left(\mathcal{E}_{m}-\delta\mathcal{E}_{\vec{\kappa}_{\nu}}\right)t},\\
I_{2} & = & \sum_{nm\vec{\kappa}_{\nu}}\Phi_{\vec{\kappa}_{\nu}}^{(nm)}e^{-i\mathcal{E}_{n}\tau}e^{-i\delta\mathcal{E}_{\vec{\kappa}_{\nu}}T}e^{i\left(\mathcal{E}_{m}-\delta\mathcal{E}_{\vec{\kappa}_{\nu}}\right)t},
\end{eqnarray*}
where $\Phi_{\vec{\kappa}_{\nu}}^{(nm)}\equiv\phi_{0}^{(n)*}\phi_{0}^{(m)}\phi_{\vec{\kappa}_{\nu}}^{(n)}\phi_{\vec{\kappa}_{\nu}}^{(m)*}$
and we have omitted the dependence on the polaron momentum $\mathbf{K}$,
i.e., $\phi_{\vec{\kappa}_{\nu}}^{(n)}(\mathbf{K})\equiv\phi_{\vec{\kappa}_{\nu}}^{(n)}$,
$\mathcal{E}_{n}\left(\mathbf{K}\right)\equiv\mathcal{E}_{n}$ and
$\delta\mathcal{E}_{\vec{\kappa}_{\nu}}(\mathbf{K})\equiv\delta\mathcal{E}_{\vec{\kappa}_{\nu}}$.
By further taking a double Fourier transformation \cite{Wang2022PRX,Jonas2003,Nardin2015}
$\mathcal{A}(\omega_{\tau},T,\omega_{t})=\iintop_{0}^{\infty}d\tau dte^{i(\omega_{\tau}+\omega_{s})\tau}\tilde{S}(\tau,T,t)e^{-i(\omega_{t}+\omega_{s})t}/\pi^{2}$,
we eventually arrive at,
\begin{equation}
\mathcal{A}=\frac{1}{2\pi^{2}}\sum_{nm\vec{\kappa}_{\nu}}\frac{\Phi_{\vec{\kappa}_{\nu}}^{(nm)}}{\omega_{\tau}^{+}-\mathcal{E}_{n}}\frac{\left[e^{-i\left(\mathcal{E}_{n}-\mathcal{E}_{m}\right)T}+e^{-i\delta\mathcal{E}_{\vec{\kappa}_{\nu}}T}\right]}{\omega_{t}^{-}-\mathcal{E}_{m}+\delta\mathcal{E}_{\vec{\kappa}_{\nu}}},\label{eq:Spectrum2DS}
\end{equation}
where $\omega_{\tau}^{+}\equiv\omega_{\tau}+i0^{+}$ and $\omega_{t}^{-}\equiv\omega_{t}-i0^{+}$,
due to their absorption and emission characteristic, respectively.
This \emph{exact} analytic expression of the 2DS of Fermi polaron
is the main result of this Letter. We aslo emphasize that our expression
of MDS can be easily generalized to Bose polaron by replacing the
multiple particle-hole excitations with Bogoliubov excitations accordingly.
A derivation of the 1DS using the same approach is given in the Supplemental
Material \cite{SM}.

\begin{figure*}
\begin{centering}
\includegraphics[width=0.9\textwidth]{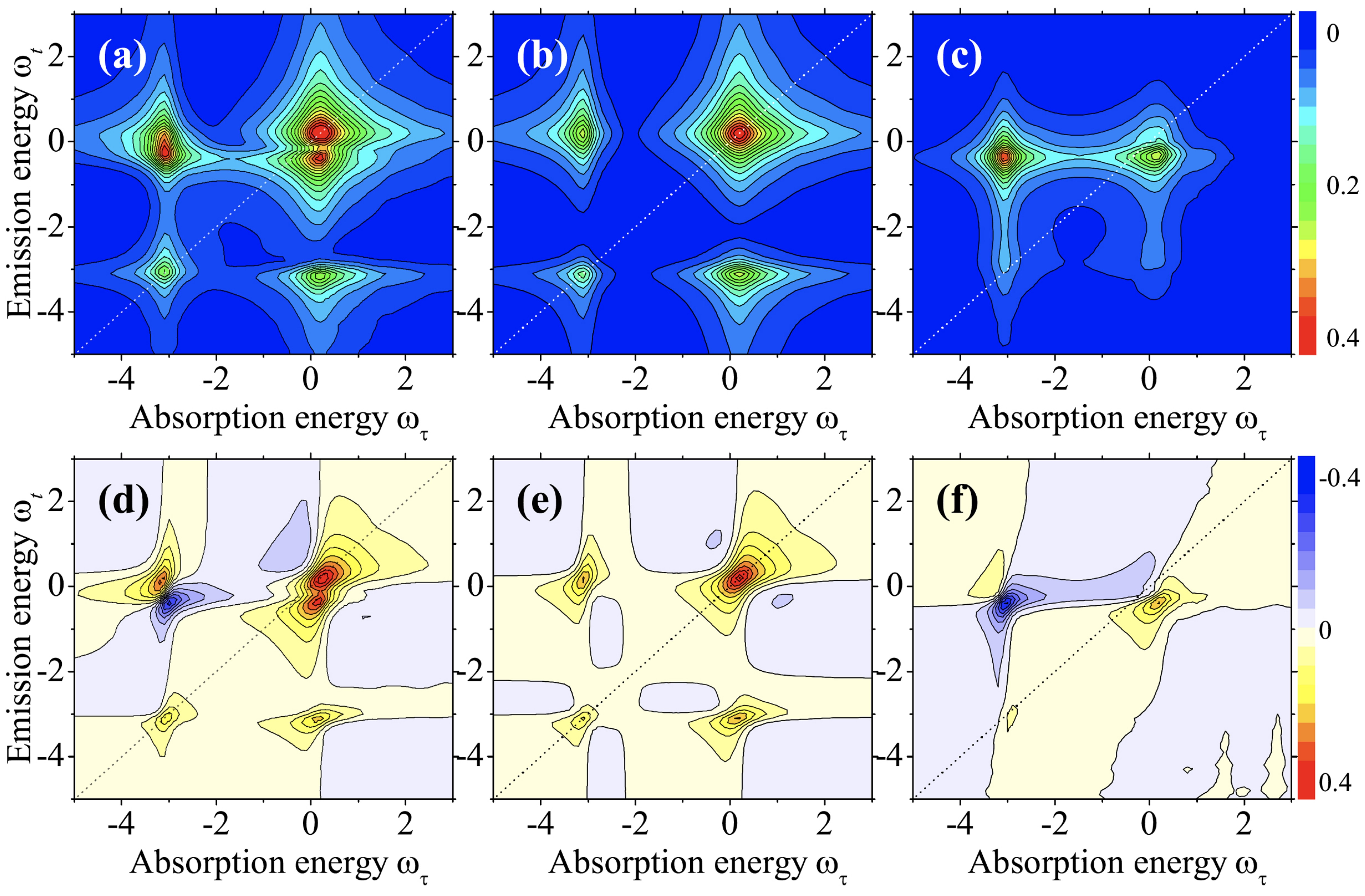}
\par\end{centering}
\centering{}\caption{Upper panel: The amplitude of the 2DS spectrum, $\left|\mathcal{A}(\omega_{\tau},T=0,\omega_{t})\right|$
(a), and its symmetric and asymmetric components, $\left|\mathcal{A}_{s}(\omega_{\tau},T=0,\omega_{t})\right|$
(b) and $\left|\mathcal{A}_{a}(\omega_{\tau},T=0,\omega_{t})\right|$
(c). Lower panel: The corresponding real part of the 2DS spectrum,
$\textrm{Re}\mathcal{A}(\omega_{\tau},T=0,\omega_{t})$ (d), $\textrm{Re}\mathcal{A}_{s}(\omega_{\tau},T=0,\omega_{t})$
(e) and $\textrm{Re}\mathcal{A}_{a}(\omega_{\tau},T=0,\omega_{t})$
(f). $\omega_{\tau}$ and $\omega_{t}$ are in units of the hopping
strength $t_{c}$, and $A(\omega_{\tau},T,\omega_{t})$ is in units
of $t_{c}^{-2}$. \label{fig:2DS_T0}}
\end{figure*}

To analyze the 2D Ramsey response, it is illustrative to truncate to one-particle-hole
excitations (i.e., the so-called Chevy ansatz \cite{Chevy2006,Cetina2016,Parish2016}),
which is known to yield quantitatively accurate attractive polaron
energy \cite{Massignan2014}. By explicitly listing the particle momentum
($\mathbf{k}_{p}$) and hole momentum ($\mathbf{k}_{h}$) in $\vec{\kappa}_{\nu=1}$
and denoting $\mathcal{\delta E}_{\vec{\kappa}_{\nu=1}}=\delta\mathcal{E}_{\mathbf{k}_{p}\mathbf{k}_{h}}=\epsilon_{\mathbf{k}_{p}}-\epsilon_{\mathbf{k}_{h}}+\epsilon_{\mathbf{K-}\mathbf{k}_{p}+\mathbf{k}_{h}}^{I}-\epsilon_{\mathbf{K}}^{I}$,
the leading order ($\mathcal{A}_{s}$) and one-particle-hole ($\mathcal{A}_{a}$)
contributions to $\mathcal{A}(\omega_{\tau},T,\omega_{t})$ can be
rewritten as,
\begin{eqnarray*}
\mathcal{A}_{s} & = & \frac{1}{2\pi^{2}}\sum_{nm}\frac{Z^{(n)}Z^{(m)}}{\omega_{\tau}^{+}-\mathcal{E}_{n}}\frac{\left[e^{-i\left(\mathcal{E}_{n}-\mathcal{E}_{m}\right)T}+1\right]}{\omega_{t}^{-}-\mathcal{E}_{m}},\\
\mathcal{A}_{a} & = & \frac{1}{2\pi^{2}}\sum_{nm\mathbf{k}_{p}\mathbf{k}_{h}}\frac{\Phi_{\mathbf{k}_{p}\mathbf{k}_{h}}^{(nm)}}{\omega_{\tau}^{+}-\mathcal{E}_{n}}\frac{\left[e^{-i\left(\mathcal{E}_{n}-\mathcal{E}_{m}\right)T}+e^{-i\delta\mathcal{E}_{\mathbf{k}_{p}\mathbf{k}_{h}}T}\right]}{\omega_{t}^{-}-\mathcal{E}_{m}+\mathcal{\delta E}_{\mathbf{k}_{p}\mathbf{k}_{h}}},
\end{eqnarray*}
where $Z^{(n)}\equiv\phi_{0}^{(n)*}\phi_{0}^{(n)}$ is the residue
of the $n$-th polaron state and $\Phi_{\mathbf{k}_{p}\mathbf{k}_{h}}^{(nm)}\equiv\phi_{0}^{(n)*}\phi_{0}^{(m)}\phi_{\vec{\kappa}_{\nu=1}}^{(n)}\phi_{\vec{\kappa}_{\nu=1}}^{(m)*}$.
It is readily seen that $\mathcal{A}_{s}(\omega_{\tau},T,\omega_{t})=\mathcal{A}_{s}^{*}(\omega_{t},T,\omega_{\tau})$
and hence the amplitude and the real part of $\mathcal{A}_{s}$ is
symmetric upon the exchange of $\omega_{\tau}$ and $\omega_{t}$.
In contrast, the one-particle-hole part $\mathcal{A}_{a}$ is not
symmetric, as a result of $\mathcal{\delta E}_{\mathbf{k}_{p}\mathbf{k}_{h}}\neq0$.

\begin{figure}[t]
\centering{}\includegraphics[width=0.5\textwidth]{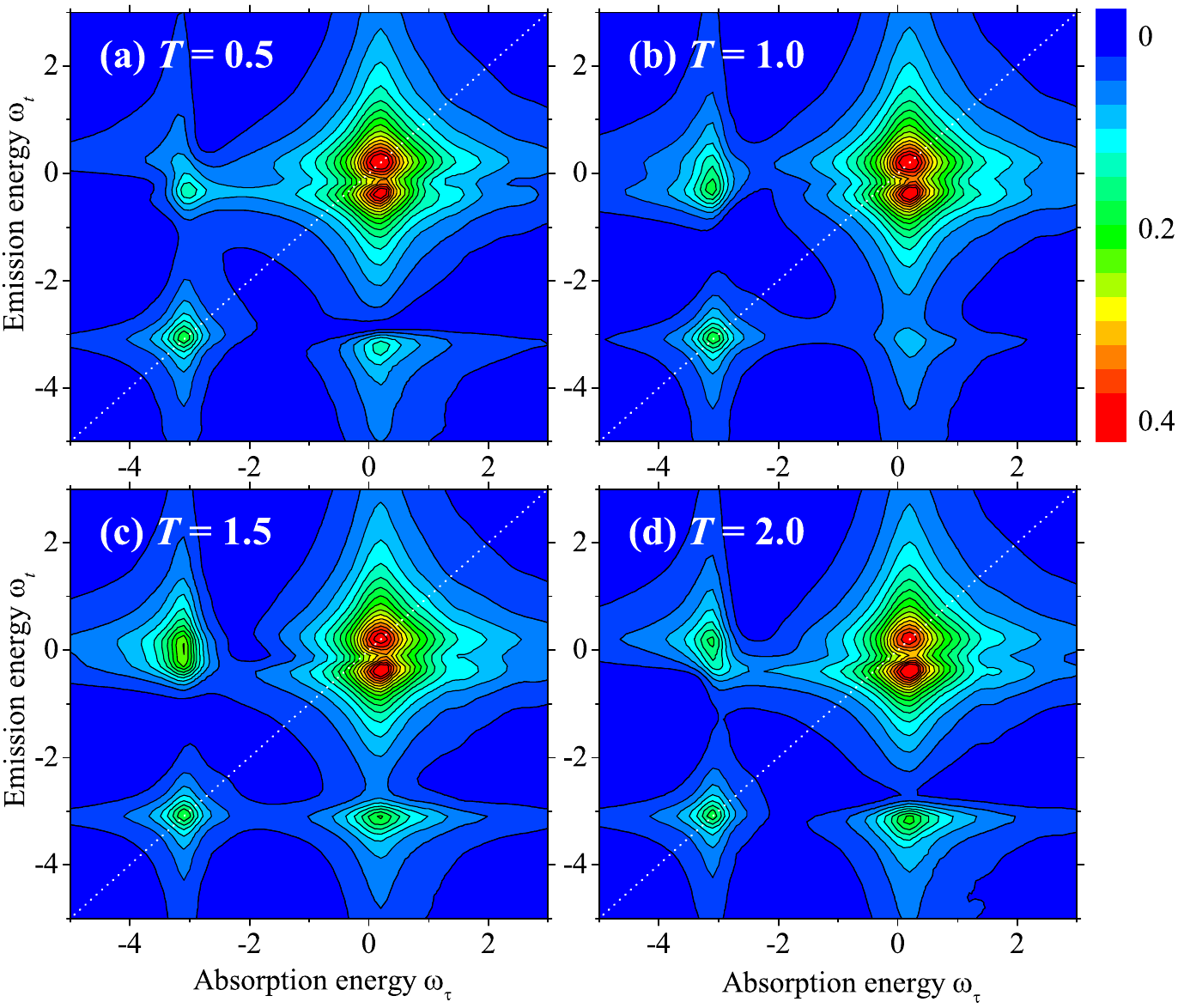}\caption{Time evolution of the amplitude of the 2DS spectrum, $\left|\mathcal{A}(\omega_{\tau},T,\omega_{t})\right|$,
under the same conditions as in Fig. \ref{fig:2DS_T0}. \label{fig:2DS_Tdep}}
\end{figure}

As a concrete example, we consider the 2DS spectrum of Fermi polarons
at the momentum $\mathbf{K}=0$ in two-dimensions, in line with the
relevant experiment on monolayer TMD materials \cite{Dey2016,Hao2016NatPhys,Hao2016NanoLett}.
For the convenience of numerical calculations, we distribute $N$
fermionic atoms on a discrete square lattice ($L\times L$) with a
hopping strength $t_{c}$. We assume the impurity has the same hopping
strength or mass as the fermionic atoms (i.e., $t_{c}=t_{d})$, so
both of them have the same dispersion relation $\epsilon_{\mathbf{k}}=-2t_{c}[\cos(k_{x})+\cos(k_{y})]=\epsilon_{\mathbf{k}}^{I}$.
We also take a relatively strong interaction $U=-8t_{c}$, which within
Chevy ansatz leads to an attractive polaron energy $E_{A}\simeq-3.08t_{c}$
with residue $Z_{A}\simeq0.24$ and repulsive polaron energy $E_{R}\simeq+0.15t_{c}$
with residue $Z_{R}\simeq1-Z_{A}$ at $N=20$ and $L=20$, as illustrated
in Fig. \ref{fig:Sketch}(a). By varying $N$ and $L$ at a filling
factor $N/L^{2}\sim0.05$, we have checked that the finite size effect
is insignificant. Throughout the work, we have used a spectral broadening
of $0.2t_{c}$, to better illustrate the 2DS spectrum.

\textbf{\textit{2DS at T=0}}\textsl{.} Figure \ref{fig:2DS_T0} presents
the amplitude and real part of $\mathcal{A}(\omega_{\tau},T=0,\omega_{t})$
and its symmetric ($\mathcal{A}_{s}$) and asymmetric ($\mathcal{A}_{a}$)
components. At zero mixing time $T=0$, $\mathcal{A}_{s}$ can be
rewritten as $\mathcal{A}_{s}(\omega_{\tau},T=0,\omega_{t})=G(\omega_{\tau})G^{*}(\omega_{t})/\pi^{2}$,
where $G(\omega)=\sum_{n}Z^{(n)}/(\omega+i0^{+}-\mathcal{E}_{n})$
is the retarded impurity Green function \cite{Massignan2014}. Therefore,
it naturally leads to the two off-diagonal crosspeaks at $(\omega_{\tau},\omega_{t})=(E_{A},E_{R})$
and $(E_{R},E_{A})$ with weight $Z_{A}Z_{R}$, in addition to the
two diagonal peaks at $(\omega_{\tau},\omega_{t})=(E_{A},E_{A})$
and $(E_{R},E_{R})$. The two crosspeaks are strongly affected by
the asymmetric one-particle-hole contribution $\mathcal{A}_{a}$,
which peaks at the upper crosspeak $(E_{R},E_{A})$ in amplitude (see
Fig. \ref{fig:2DS_T0}(c)). As a result, the two crosspeak become
highly asymmetric, as shown in Fig. \ref{fig:2DS_T0}(a). $\mathcal{A}_{a}$
is also significant near the diagonal peak at the repulsive energy
$(E_{R},E_{R})$, forming a side peak slightly below it. 

\textbf{\textit{Quantum oscillations}}. The existence of the two highly
asymmetric crosspeaks in 2DS spectrum is a strong evidence of the
quantum coherence between attractive and repulsive polarons. Further
smoking-gun confirmation can be provided by quantum beats between
the crosspeaks at different mixing time $T$, as reported in Fig.
\ref{fig:2DS_Tdep}. From the expression of $\mathcal{A}(\omega_{\tau},T,\omega_{t})$
in Eq. (\ref{eq:Spectrum2DS}), it is readily understood that these
beats are caused by the term $e^{-i\left(\mathcal{E}_{n}-\mathcal{E}_{m}\right)T}$,
which leads to an oscillation with periodicity $2\pi/\left|E_{A}-E_{R}\right|$
and decay rate $\Gamma_{R}$, where $\Gamma_{R}$ is the decay rate
of the repulsive polaron \cite{Massignan2014,Massignan2011}. This
term does not affect the two diagonal peaks, so the 2D spectrum near
them is essentially independent on the mixing time $T$, as can be
seen from Fig. \ref{fig:2DS_Tdep}.

To better characterize the quantum oscillations, we study $\textrm{Re}\mathcal{A}(\omega_{\tau},T,\omega_{t})$
at the two crosspeaks, respectively
labelled as AR (lower crosspeak) and RA (higher crosspeak). As shown in Fig. \ref{fig:Re2DS_CrosspeakTdep} of Supplemental Material \cite{SM}, despite the same periodicity, interestingly, the two oscillations at AR
and RA crosspeaks are not synchronized. The different phases of the
two oscillations could be due to the term $e^{-i\delta\mathcal{E}_{\mathbf{k}_{p}\mathbf{k}_{h}}T}$
in the asymmetric one-particle-hole component $\mathcal{A}_{a}$.
Indeed, we find
$\textrm{Re}\mathcal{A}_{a}$ behaves very different at AR and RA, in sharp contrast to $\textrm{Re}\mathcal{A}_{s}$,
which gives the exactly same value at the two crosspeaks.

\textbf{\textit{Relevance to 2DCS}}\textsl{.} We now compare our theoretical
results to the recent 2DCS experiment on Fermi polarons consisting
of excitons and trions in monolayer TMD materials \cite{Hao2016NanoLett}.
Although the ways for implementing 2D spectroscopy are different,
our simulated 2DS spectrum for Fermi polarons reproduce the key experimental
observations \cite{Hao2016NanoLett}, such as the appearance of the
two off-diagonal crosspeaks and their quantum beats as a function
of the mixing time $T$. Thereby, in principle our microscopic many-body
calculation presents an exciting full ab initio account of the 2DCS
spectroscopy of mobile polaron, which has never been achieved, to the best of our knowledge.

\textbf{\textit{Conclusions}}\textsl{.} We have predicted that quantum
beats between the two off-diagonal crosspeaks in the recently proposed
two-dimensional Ramsey spectroscopy \cite{Wang2022PRX} for Fermi
polarons are ideally suited to unveil the quantum coherence between
the attractive and repulsive polaron branches. Our theoretical results
are able to capture the key features of a recent experiment on Fermi
polaron-excitons in atomically thin transition metal dichalcogenides
\cite{Hao2016NanoLett} and could be quantitatively verified in highly
controllable cold-atom experiments in the near future.
\begin{acknowledgments}
This research was supported by the Australian Research Council's (ARC)
Discovery Program, Grants No. DE180100592 and No. DP190100815 (J.W.),
and Grant No. DP180102018 (X.-J.L).
\end{acknowledgments}

\setcounter{figure}{0}
\renewcommand{\figurename}{Fig.}
\renewcommand{\thefigure}{S\arabic{figure}}
\setcounter{equation}{0}
\renewcommand{\theequation}{S\arabic{equation}}

\section{Supplemental Materials}
\subsection{Derivation of 1D spectroscopy}
In this Supplemental Material, we give a derivation of the conventional
1D Ramsey response and spectral function. In 1D Ramsey scheme, only
one $\pi/2$ rf pulse is applied before the last detection pulse at
time $\tau$, which gives the final state

\[
\left|\psi_{f}\right\rangle =\mathcal{U}\left(\tau\right)\frac{d_{\mathbf{K}\uparrow}^{\dagger}+d_{\mathbf{K}\downarrow}^{\dagger}}{\sqrt{2}}\otimes\left|\textrm{FS}\right\rangle .
\]
The Ramsey response can be obtained by measuring $\sigma_{-}=\sigma_{x}-i\sigma_{y}=2\hat{s}_{-}$,
\begin{equation}
\tilde{S}_{a}(\tau)=2\left\langle \psi_{f}\left|\hat{s}_{-}\right|\psi_{f}\right\rangle =\left\langle \textrm{FS}\right|d_{\mathbf{K}\downarrow}\hat{I}_{a}d_{\mathbf{K}\uparrow}^{\dagger}\left|\textrm{FS}\right\rangle ,
\end{equation}
where the pathway is given by
\begin{equation}
\hat{I}_{a}\equiv e^{i\mathcal{H}_{\downarrow}\tau}\hat{s}_{-}e^{-i\mathcal{H}_{\uparrow}\tau}.
\end{equation}
Measurement of $\sigma_{-}$ is in consistent with the 2DS measurement
of $-\sigma_{+}$ with three $\pi/2$ pulses and $t=T=0$. The additional
two instantaneous $\pi/2$ pulses can be recognized as a unitary transformation
$U_{\pi}=i\sigma_{y}$, which gives $-U_{\pi}^{\dagger}\sigma_{+}U_{\pi}=\sigma_{-}$.

We expand the time evolution operators as ($t'=\tau,T,t$),
\begin{eqnarray*}
e^{\pm i\mathcal{H}_{\uparrow}t'} & = & \sum_{n\mathbf{k}}e^{\pm iE_{\uparrow}^{(n)}(\mathbf{k})t'}\left|n;\mathbf{k}\right\rangle \left\langle n;\mathbf{k}\right|,\\
e^{\pm i\mathcal{H}_{\downarrow}t'} & = & \sum_{\vec{\kappa}_{\nu}\mathbf{k}}e^{\pm iE_{\downarrow}^{(\vec{\kappa}_{\nu})}(\mathbf{k})t'}\left|\vec{\kappa}_{\nu};\mathbf{k}\right\rangle \left\langle \vec{\kappa}_{\nu};\mathbf{k}\right|,
\end{eqnarray*}
with polaron states (with index $n$)
\begin{align*}
\left|n;\mathbf{k}\right\rangle  & =\sum_{\vec{\kappa}_{\nu}}\phi_{\vec{\kappa}_{\nu}}^{(n)}\left(\mathbf{k}\right)d_{\mathbf{k}-\mathbf{k}_{\vec{\kappa}_{\nu}}\uparrow}^{\dagger}\left|\vec{\kappa}_{\nu}\right\rangle ,
\end{align*}
where $\left|\vec{\kappa}_{\nu}\right\rangle =\prod_{i=1}^{\nu}c_{\mathbf{k}_{p}^{(i)}}^{\dagger}\prod_{i=1}^{\nu}c_{\mathbf{k}_{h}^{(i)}}\left|\textrm{FS}\right\rangle $
denotes $\nu$ particle-hole pairs excitations on top of a Fermi sea,
$\vec{\kappa}_{\nu}\equiv\{\mathbf{k}_{p}^{(1)},\mathbf{k}_{p}^{(2)},\cdots,\mathbf{k}_{p}^{(\nu)};\mathbf{k}_{h}^{(1)},\mathbf{k}_{h}^{(2)},\cdots,\mathbf{k}_{h}^{(\nu)}\}$
is a collective notation for the $\nu$ particle momenta ($\mathbf{k}_{p}^{(i)}>k_{F}$)
and hole momenta ($\mathbf{k}_{h}^{(i)}<k_{F}$), and therefore the
total momentum and energy of the particle-hole excitations are given
by $\mathbf{k}_{\vec{\kappa}_{\nu}}\equiv\sum_{i=1}^{\nu}[\mathbf{k}_{p}^{(i)}-\mathbf{k}_{h}^{(i)}]$
and $\epsilon_{\vec{\kappa}_{\nu}}=\sum_{i=1}^{\nu}[\epsilon_{\mathbf{k}_{p}^{(i)}}-\epsilon_{\mathbf{k}_{h}^{(i)}}]$,
respectively. The energy of the ($n$-th) polaron state can be denoted
as, $\mathcal{E}_{n}(\mathbf{k})=E_{\uparrow}^{(n)}(\mathbf{k})-(\epsilon_{\mathbf{k}}^{I}+\omega_{s}+E_{\textrm{FS}})$,
after the subtraction of the impurity energy ($\epsilon_{\mathbf{k}}^{I}+\omega_{s}$)
and the energy of the background Fermi sea ($E_{\text{FS}}$). On
the other hand, the many-body eigenstates in the case of the spin-down
impurity are much simpler and can be directly characterized by $\vec{\kappa}_{\nu}$,
i.e., $\left|\vec{\kappa}_{\nu};\mathbf{k}\right\rangle =d_{\mathbf{k}-\mathbf{k}_{\vec{\kappa}_{\nu}}\downarrow}^{\dagger}\left|\vec{\kappa}_{\nu}\right\rangle $.
The corresponding energy is given by, $\delta\mathcal{E}_{\vec{\kappa}_{\nu}}(\mathbf{k})=E_{\downarrow}^{(\vec{\kappa}_{\nu})}(\mathbf{k})-(\epsilon_{\mathbf{k}}^{I}+E_{\textrm{FS}})=\epsilon_{\vec{\kappa}_{\nu}}+\epsilon_{\mathbf{\mathbf{k}-\mathbf{k}_{\vec{\kappa}_{\nu}}}}^{I}-\epsilon_{\mathbf{k}}^{I}$,
which is a summation of the recoil energy of the impurity and the
Fermi sea.

Inserting the expansion of time evolution operators into $\tilde{S}_{A}(\tau)$
gives the 1D Ramsey response
\begin{equation}
\tilde{S}_{A}(\tau)=\sum_{n}e^{-i\mathcal{E}_{n}(\mathbf{k})t}\phi_{0}^{(n)*}\phi_{0}^{(n)}\equiv\sum_{n}e^{-i\mathcal{E}_{n}(\mathbf{k})t}Z^{(n)},
\end{equation}
which is related with the spectral function $\mathcal{A}_{A}(\omega_{\tau})$
by a Fourie transformation 
\begin{equation}
\mathcal{A}_{A}(\omega_{\tau})=\frac{1}{\pi}\int_{0}^{\infty}d\tau e^{i(\omega_{\tau}+\omega_{s})\tau}\tilde{S}_{A}(\tau)=\frac{1}{\pi}\sum_{n}\frac{Z^{(n)}}{\omega_{\tau}^{+}-\mathcal{E}_{n}},
\end{equation}
where we omit the dependence of $\mathbf{K}$ in $Z^{(n)}(\mathbf{K})$
and $\mathcal{E}_{n}(\mathbf{K})$ for simplicity of notation. These
expressions is in consistent with previous studies \citep{Massignan2014,Schmidt2018}.

\subsection{Quantum oscillations at the crosspeaks}
\begin{figure}[h]
\centering{}\includegraphics[width=0.96\columnwidth]{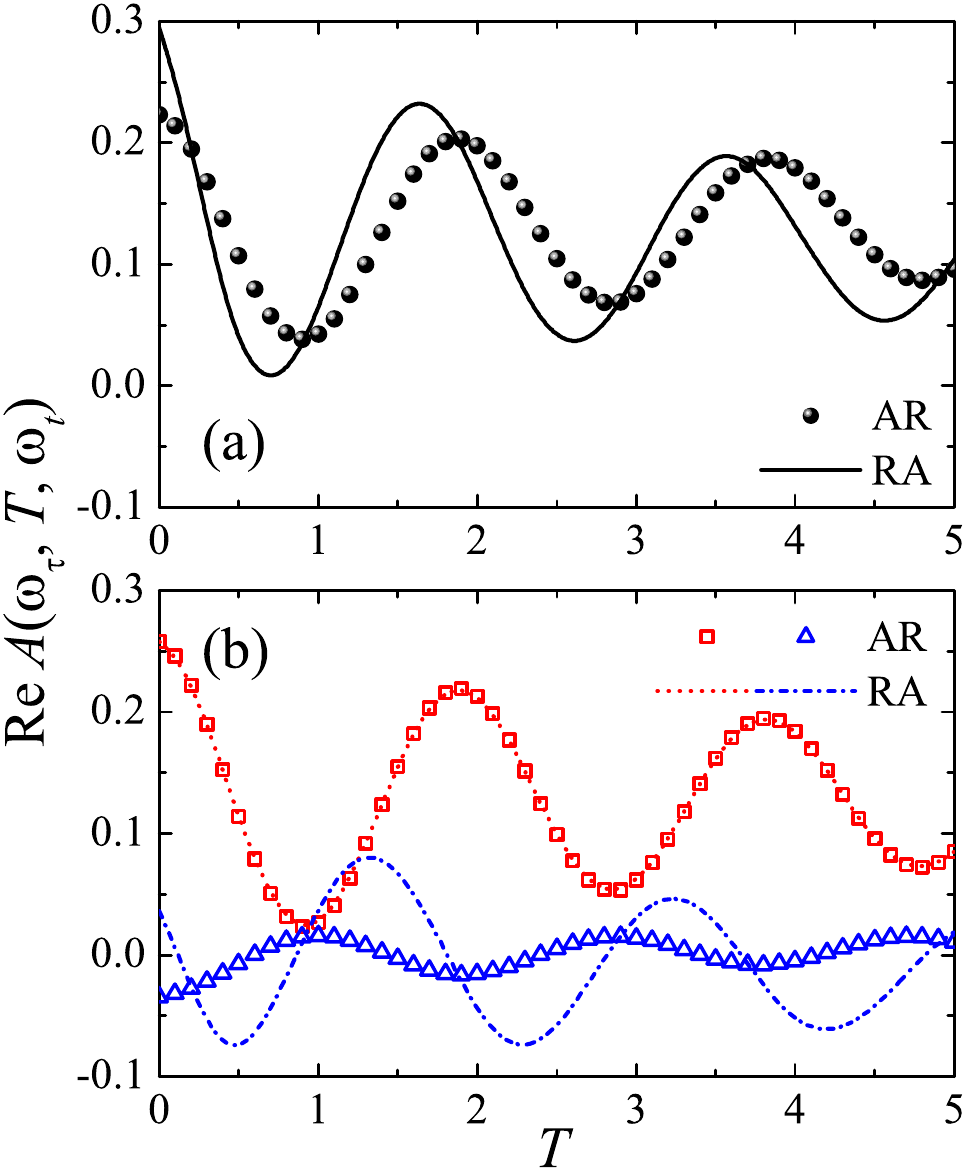}\caption{(a) Time-dependent real part of the 2DS $\textrm{Re}\mathcal{A}(\omega_{\tau},T,\omega_{t})$
at the lower crosspeak (AR, black circles) and the higher crosspeaks
(RA, black solid line). (b) $\textrm{Re}\mathcal{A}_{s}(\omega_{\tau},T,\omega_{t})$
(red dashed line and squares) and $\textrm{Re}\mathcal{A}_{a}(\omega_{\tau},T,\omega_{t})$
(blue dot-dashed line and triangles) at the lower crosspeak (AR, symbols)
and the higher crosspeaks (RA, lines). Other parameters are the same
as in Fig. 2 in the main text. \label{fig:Re2DS_CrosspeakTdep}}
\end{figure}

To illustrate the quantum oscillations at the crosspeaks in details, we show $\textrm{Re}\mathcal{A}(\omega_{\tau},T,\omega_{t})$
at the two crosspeaks, respectively labelled as AR (lower crosspeak) and RA (higher crosspeak) as a function of $T$ in Fig. (\ref{fig:Re2DS_CrosspeakTdep}). As shown in Fig. \ref{fig:Re2DS_CrosspeakTdep}(b)
$\textrm{Re}\mathcal{A}_{a}$ behave very different at AR and RA (see the blue triangles and dot-dashed line). This is in sharp contrast to $\textrm{Re}\mathcal{A}_{s}$, which gives the exactly same value at the two crosspeaks (see the overlapping red squares and dotted line).

\end{document}